\PassOptionsToPackage{dvipsnames}{xcolor}	
\documentclass[12pt,letter]{article}

\usepackage[T1]{fontenc}
\usepackage{lmodern}
\usepackage{setspace}
\usepackage{etoolbox}
\makeatletter
\patchcmd{\appendix}{\@Alph}{\@Roman}{}{}
\makeatother

\usepackage{amsmath}
\usepackage{amssymb}
\usepackage{amsthm}
\usepackage{mathtools}
\usepackage{mathrsfs}
\usepackage{breqn}
\usepackage{bigints}

\usepackage[margin=1 in]{geometry}
\usepackage{enumerate}
\usepackage{enumitem} 
\usepackage{xcolor}
\usepackage[colorlinks=true, urlcolor=blue, linkcolor=black, citecolor=black]{hyperref}
\usepackage{float}
\usepackage{indentfirst}
\usepackage{caption}
\usepackage{fancyhdr}
\usepackage{tcolorbox}
\usepackage{multirow,array}
\usepackage{bbm}

\usepackage{tikz}
\usetikzlibrary{decorations.pathreplacing}
\usepackage{mathtools}
\usetikzlibrary{positioning}
\usepackage{graphicx}

\renewcommand{\epsilon}{\varepsilon}

\newtheorem{proposition}{Proposition}

\newtheorem*{statement*}{Result}

\theoremstyle{definition}

\DeclareMathOperator*{\argmax}{\arg\!\max}

\DeclareTextFontCommand{\emph}{\slshape}

\usepackage{natbib}
\nocite{*}

\title{Artificial or Human Intelligence?}

\author{Eric Gao\thanks{Department of Economics, MIT. ericgao@mit.edu. I would like to thank Joshua Gross, Hongbin Li, Daniel Luo, Eric Tang, and Bryant Xia for helpful comments and discussions.}}

\date{\today}

\begin{document}
	
\maketitle

\begin{abstract}
Artificial intelligence (AI) tools such as large language models (LLMs) are already altering student learning. Unlike previous technologies, LLMs can independently solve problems regardless of student understanding, yet are not always accurate (due to hallucination) and face sharp performance cutoffs (due to emergence). Access to these tools significantly alters a student's incentives to learn, potentially decreasing the sum knowledge of humans and AI. Additionally, the marginal benefit of learning changes depending on which side of the AI frontier a human is on, creating a discontinuous gap between those that know more than or less than AI. This contrasts with downstream models of AI's impact on the labor force which assume continuous ability. Finally, increasing the portion of assignments where AI cannot be used can counteract student mis-specification about AI accuracy, preventing underinvestment. A better understanding of how AI impacts learning and student incentives is crucial for educators to adapt to this new technology.
\end{abstract}

	
\newpage

\onehalfspacing

\section{Introduction}

Many open questions remain about how artificial intelligence (AI) will impact society. Two of the most pressing questions are whether or not AI will augment or replace human intelligence and whether AI leads to greater or lesser inequality. While there has been much recent work investigating these questions both empirically and theoretically (\cite{brynjolfsson_2025_canaries}, \cite{brynjolfsson_2025_generative}; \cite{ide_2025_artificial}), much of this work has focused on the impacts of AI in the workforce. Such work is unsatisfactory for two reasons. First, such analyses start off by assuming some exogenously determined baseline level of ability for each worker. Second, tasks in the workforce are already pre-defined and workers have little opportunity to improve and develop their underlying ability.

Instead, ability is determined at an earlier stage: during a future worker's education. At many educational institutions across all levels, AI (and large language models in particular) has already begun to play a significant role. It has led to many positive impacts, such as helping students learn new concepts by providing personalized and immediate feedback. However, it has also led to concerns about students relying on AI too much by using it to complete assignments and exams, leading to less learning and skill development \citep{vieriu_2025_the, kaledio_2024_the, zhai_2024_the}. Just like how AI is a transformative and unprecedented technology in the workforce due to its ability to autonomously solve problems, it is also a transformative and unprecedented technology in education for the same reasons. Unlike previous educational aids that require some degree of human understanding and input to use for solving specific problems, AI models can ``one shot'' problems with minimal human input (at varying degrees of accuracy), taking students out of the equation. As Anthropic, the developer of Claude (a leading LLM) reports, ``nearly half ($\sim$47\%) of student-AI conversations were Direct—that is, seeking answers or content with minimal engagement'' from a sample of a million conversations verified students had with Claude \citep{a_2025_anthropic}. In particular, they documented

\begin{quote}
Concerning Direct conversation examples including:

- Provide answers to machine learning multiple-choice questions \\
- Provide direct answers to English language test questions \\
- Rewrite marketing and business texts to avoid plagiarism detection.
\end{quote}

Unfortunately, directly regulating AI usage is difficult. Institutionally, it is difficult to ``prove'' that a student has indeed used AI on an assignment to officially take disciplinary action. AI is becoming increasingly difficult to detect as models improve and as students learn to better prompt AI models to generate more human-like responses. Furthermore, just like how restricting usage of calculators or other tools is not representative of the real world, forcing students to not use AI on assignments or exams may similarly lead to worse learning outcomes or under-utilization of a powerful tool. As such, understanding student incentives to learn and invest in their own ability change when they have access to AI tools that can both autonomously solve (some) problems and help students learn is crucial to understand student outcomes and how educators can best adapt to this new technology.

To answer these questions, I study a model incorporating two stylized facts about AI: (1) it exhibits a sharp drop in performance after a certain point (similar to \cite{ide_2025_artificial} modeling AI at a certain level of knowledge) and (2) even among problems it can solve, it sometimes hallucinates. Students have access to this technology and decide how much costly investment in their own ability to undertake. Each student has some type that governs how costly it is for them to learn, where higher types face lower costs (and thus choose higher levels of ability ceteris paribus). Learning what the AI already knows produces lower marginal benefits (as humans are only useful in case AI hallucinates) while learning beyond the scope of what AI has a larger marginal benefit (as AI is unable to solve these problems independently) but requires more investment. 

This simple, stylized model makes several predictions that unify empirical findings on how AI has impacted both labor and learning. First, given any exogenous distribution of fixed individual abilities, the introduction of AI benefits those who are the least skilled. The higher ability an individual is, the more problems they can solve on their own, which means that more of their capabilities overlap with the AI, diminishing its benefit. For individuals with ability beyond AI, introduction of AI does not do anything to benefit them (and can only hurt their accuracy if they rely on it to solve any problems). On the other hand, if instead of starting with ability each agent had some type that governs their cost of learning, there exists some threshold type for which humans below this type are less knowledgeable than AI and those above this type are more knowledgeable. This threshold type is characterized by indifference: Students of this type are indifferent between having less or more knowledge than AI. Furthermore, there is a discontinuous gap in ability at this type, as humans on either side face different marginal benefits to learning. I say that students who are less knowledgeable than AI use it as a \textit{solver}, since AI is used to independently solve problems. On the other hand, students who are more knowledgeable than AI use it as a \textit{helper}.

Next, I characterize how changes in AI technology shifts student incentives to learn. Increasing the accuracy of AI (i.e. reducing hallucinations) has two implications. First, it decreases the marginal benefit to learning for students who use AI as a solver but does not impact the marginal benefit to learning for students who use AI as a helper, leading to less investment in ability for those that have a lower type. Second, it pushes the threshold type upwards, as the benefits to using AI as a solver increase relative to using AI as a helper at the previous threshold type. On the other hand, increasing the difficulty of problems AI can solve does not change the marginal benefit of any agent, but still increases the threshold type by once again benefiting a student of the threshold type who uses AI as a solver, breaking indifference. 

However, things change when we consider AI reducing costs to learn. If AI is complementary with learning more difficult subjects (i.e. it reduces the cost to learn more at higher levels of ability), then students generally choose higher ability and the threshold type decreases. The magnitude of how much AI needs to augment the learning process for it to not simply replace human intelligence depends on how advanced AI is.

Finally, I consider misspecified students who over-estimate how accurate AI is. Such students under-invest in their own ability compared to correctly specified students. However, instructors have flexibility in how they design assignments and exams to remedy such misspecification. By putting some weight on assignments that do not permit the usage of AI (such as in-person tests or timed writings), instructors can incentivize misspecified students to invest more in their own ability. The optimal weight on assignments that do not permit the usage of AI increases as the degree of student misspecification increases.

The remainder of the paper is structured as follows. The next section discusses related literature and details of how the conclusions of this model relate. Section 3 discusses technical background on AI hallucinations and the emergence, the two stylized building blocks of the model. Section 4 presents the model. Section 5 characterizes comparative statics. Section 6 discusses misspecified students and how educators can optimally design assignments and exams to mitigate under-investment in ability. Section 7 concludes.

\section{Related Literature}

The closest paper is \cite{ide_2025_artificial}, which studies how AI impacts labor markets. In their model, worker ability is exogenously fixed over $[0, 1]$ (which corresponds to a set of problem difficulties) and AI can solve problems up to some threshold. When a worker is faced with a problem they cannot solve, they can pass the problem onto a (human or AI) solver who has a higher ability. With the introduction of AI, some humans become solvers (solving problems beyond the scope of AI) and others become workers (solving low-difficulty problems independently and identifying which problems are high-difficulty to pass onto human or AI solvers). If AI is autonomous, high-knowledge workers are benefited more than low-knowledge workers. If AI is not autonomous, low-knowledge workers are benefited more than high-knowledge workers. One of the assumptions of their model was that the distribution of worker knowledge is continuous. However, as this paper suggests, allowing for endogenous costly investment in ability leads to a discontinuity in the distribution of worker ability, precisely around the level of knowledge AI has. In \cite{ide_2025_artificial}, workers with knowledge further awy from AI's knowledge (in either direction) are benefited more, suggesting pooling towards either extreme (low or high knowledge) if it were endogenous.

Beyond this early theoretic work, many empirical papers have also studied the impact AI has on the workforce. \cite{brynjolfsson_2025_canaries} uses payroll data to study the impact of AI on the labor market. The first major finding is that employment opportunities for young workers in fields most exposed to AI have been harmed the most. Young workers do not have the tacit knowledge that older workers have; it is precisely this tacit knowledge that is difficult for AI to replicate. Instead, young workers mainly only have codified knowledge, which is most exposed to AI replacement. However, these aggregate statistics do not paint a complete picture of the situation. Even in the domain of codified knowledge, \cite{ide_2025_artificial} and this paper both suggest that whether human knowledge is above or below AI knowledge plays a crucial role in determining how AI impacts humans. One area where humans with codified knowledge are still experiencing rapid job growth is that of AI research, where AI companies and startups are aggressively hiring junior employees, often right out of college \citep{bindley_2025_these}. 

Next, \cite{brynjolfsson_2025_generative} studies the impact of AI adoption among customer-support agents in the workforce. \cite{noy_2023_experimental} and \cite{doshi_2024_generative} experimentally analyze how LLM assistance impacts writing-based tasks. \cite{caplin_2024_the} is also experimental, investigating how AI aids benefit human decision-makers using the task of predicting whether the subject of a picture was above 21 years of age or not. All studies found that AI helped reduce inequality by benefiting lower-ability workers more than higher-ability workers. However, these works also exposed potential downsides of AI adoption. \cite{brynjolfsson_2025_generative} finds that for high-skilled workers, introduction of AI decreases their resolution rate (while leading to small reductions in chat times). This corresponds to a human with ability greater than AI using it to solve problems it can do without intervening. Furthermore, AI tools are trained on high-skilled workers, leading to less novel training data if these workers begin to rely on AI. \cite{doshi_2024_generative} has writer creativity as their desiderata: When assisted with AI, each individual writer's stories became more creative, but when viewed as a population, stories became more creative in the same direction, lessening overall group creativity. Finally, \cite{caplin_2024_the} cautions that human decision-makes who are miscalibrated about their confidence in their own ability accrue less benefits from AI due to relying on AI assistance too little (if they are overconfident) or too much (if they are under-confident). This paper considers a different form of miscalibration: Students may over-estimate how accurate AI is, leading to under-investment in their own ability. 

Finally, there the literature studying how AI impacts learning and education. While most prior work has been qualitative, there has been some recent empirical work. \cite{bastani_2024_generative} conducts a field experiment introducing GPT-4 to high-school math classes. Some students were not provided with the LLM, others were provided full access, and a final group was only provided with ``GPT Tutor,'' which has built-in safeguards to preserve learning (i.e. refusing to directly give solutions). Compared to the group with no GPT access, students with access to GPT saw a 48\% increase in performance on practice problems, but a 17\% decrease in performance on an exam where the LLM was taken away. On the other hand, students with access to GPT Tutor saw a 127\% increase in performance on practice problems and no significant change in exam performance. One drawback of this study was that GPT tutor was provided solutions to all practice problems (while GPT alone made errors on around half of the problems), potentially explaining why students with GPT tutor performed significantly better on practice problems. Looking at how students behave when GPT base is wrong suggests that students are, in fact, simply copying LLM output instead of understanding the problem: GPT errors significantly impact practice problem performance (even arithmetic ones that students should easily be able to catch) but do not impact exam performance. Furthermore, students engaged significantly more superficially with GPT compared to GPT tutor.

\cite{LehmannCorneliusSting2024} studies similar questions in the lab, having students complete a python programming course with or without access to GPT. Findings are largely similar: Students asked for solutions vastly more than explanations, especially when students were able to directly copy-paste GPT output. In that case, ``42\% of messages asking for a solution were sent without a single attempt to solve the corresponding question.'' GPT access made students faster at completing individual problems, increasing the number of topics students covered but decreased their understanding of any individual topic. Additionally, this paper finds evidence that LLM access may increase inequality between students: Participants that scored higher on a pre-test benefited more from LLM access than those that scored lower. Finally, students with access to GPT were more likely to over-estimate their own ability, another potential pitfall to consider when evaluating LLMs as an educational tool.

In a different setting, \cite{riedl_2024_effects} studies how the introduction of AI-based chess engines impacts player learning. Contrary to findings of AI decreasing inequality in labor markets, access to the tool also led to increased inequality by helping strong players more than weak ones. However, the mechanism they find is different from the one identified in this paper. The authors first establishes that AI feedback is only helpful after a loss, whereas seeking feedback after a win (as a pat on the back) is harmful. Then, stronger players seek feedback more often after losses, leading to more learning compared to weaker players. Similar to \cite{doshi_2024_generative}, players' choices of which opening to play also became less diverse after the introduction of AI.

\section{Technical Background}

\subsection{Hallucinations}

Large language models are fundamentally probabilistic. Upon receiving a sequence of input text, an LLM breaks it down into tokens (words or sub-words) and predicts the next token based on previous tokens. Individual tokens are embedded into a high-dimensional space and are then passed through multiple layers of attention (to learn context) and feed-forward neural networks. The output of the final layer is then transformed into a probability distribution over the entire vocabulary of possible tokens using a softmax function, and the next token is sampled from this distribution. This process is repeated until a stopping criterion is met (such as generating a special end-of-sequence token or reaching a maximum length). When training LLMs, the model seeks to minimize divergence between predicted token probabilities and the actual tokens in the training data. This means that the model is simply incentivized to assign higher probabilities to tokens that are more likely to appear in the training data given the preceding context.

As such, LLMs are (1) not trained to be ``correct'' but rather to produce text that is statistically likely given the input and (2) inherently probabilistic, meaning that even if the model produces a distribution of tokens that places a high probability on the correct token, there is still a chance that an incorrect token is sampled. As such, even in a perfect world where data quality concerns, training costs, and model architecture are not an issue, LLMs will still hallucinate with some probability. In fact, there has been recent research showing that there cannot exist any LLM that is completely free of hallucinations, even in these perfect worlds \citep{xu_2024_hallucination, banerjee_2024_llms}. 

In practice, AI hallucination rates depend significantly on the task and benchmark chosen. Even when evaluated on the simple task of summarizing an article that is provided to the LLM, the best models  \href{https://github.com/vectara/hallucination-leaderboard?tab=readme-ov-file}{hallucianate around one percent} of the time. When moving to the more complicated benchmark of asking LLMs questions about a set of news articles, \href{https://research.aimultiple.com/ai-hallucination/}{hallucination rates jumped to 15\% for the best models}. When using AI in the field, outside of these controlled benchmarks, hallucination rates are often much higher. In \cite{bastani_2024_generative}, GPT-4 made errors on around half of the high-school math problems it was asked to solve. When answering legal questions, hallucination rates can jump to over 70\% \citep{emslie_2024_llm}.

\subsection{Emergence}

Another core component of LLM capabilities is the phenomenon of emergence. As LLM model size, training data, and computational resources have been scaled up, they have exhibited sudden and unpredictable improvements in performance on certain tasks. These improvements go beyond the scope of simple scaling laws: Simply taking the performance of small models and scaling them appropriately would not have predicted emergent capabilities. Instead, they arise from complex (and sometimes not fully understood) interactions between the size of model parameters and its training data.

A wide range of abilities have been documented as emergent, from addition to unscrambling words to more general question and answer tasks \citep{wei_2022_emergent}. There are structural reasons for this phenomenon. For example, if a task requires $k$ levels of reasoning, then a model would require at least $k$ hidden layers in its neural network to perform the task. If a model has fewer than $k$ layers, it cannot perform the task at all (it performs ``no better than random''), leading to a sharp jump in performance when models become sufficiently large. Another example is that of addition: Internal layers of the model process addition digit by digit (called Implicit Discrete State Representations) but such internal steps are not displayed in the final output. Once again, models need a sufficiently high number of parameters for this to be possible \citep{berti2025emergent}. 

One alternative view of emergence is that instead of LLMs exhibiting discontinuous jumps in performance, discontinuities come from benchmarks being discontinuous instead \citep{schaeffer_2023_are}. If we looked at an alternative measure of accuracy instead of a binary measure of whether or not output is perfectly correct, growth scales more continuously. However, for the purposes of how students utilize LLMs, they are primarily concerned with whether or not the LLM can solve a problem correctly instead of alternative measures of how many tokens off a LLM was.

\section{Model}

There are problems of difficulties (uniformly distributed) over the interval $[0, 1]$. AI can solve problems up to difficulty $d \in [0, 1]$ with probability $p \in [0, 1]$ and no problems of difficulty greater than $d$. Having $p < 1$ captures the fact that AI sometimes hallucinates even on problems it is capable of solving, while a sharp cutoff at $d$ captures AI abilities being emergent. Both $p, d$ are exogenously determined by the state of AI technology; how student behavior changes as AI technology varies is of interest. Changes in $p$ could correspond to recent advances in chain-of-thought reasoning, reinforcement learning, or better model architectures, while changes in $d$ could correspond to larger models trained on more data.

Additionally, there is a human (student) of type $t \in [0, 1]$ can invest cost $c(a, t| d, p)$ to acquire ability level $a \in [0, 1]$ (so changes in $d, p$ also influence the cost of learning). Suppose $c$ satisfies:
\begin{enumerate}
    \item All second-order derivatives exist.
    \item Decreasing differences in $a, t$ for any $d, p$.
\end{enumerate} 
A human with ability level $a$ can solve problems up to difficulty $a$ with one hundred percent accuracy and no problems of difficulty larger than $a$. However, humans can also use AI to help them solve problems. The set of problems a human can solve with AI assistance is simply the union of the problems they can solve and the problems the AI can solve. If a human chooses $a < d$ then we say the human uses AI as a solver (as the AI is independently solving some problems that are beyond the scope of the human) while if $a > d$ then we say the human uses AI as a helper (as the human able to solve all the problems the AI can solve but still uses the AI to learn at a lower cost).

For example, the following diagram represents the sets of problems that an AI with $d = 3/4, p = 1/2$ and a human with $a = 1/2$ can solve:

\begin{figure}[h]
    \centering
    \begin{tikzpicture}

    \draw (0,0) -- (4.2,0);
    \draw (0,0) -- (0,4.2);

    \draw (0,0.1) -- (0,-0.1);
    \node[below] at (0,-0.15) {0};
    \draw (4,0.1) -- (4,-0.1);
    \node[below] at (4,-0.15) {1};

    \draw (0,0) -- (-0.1,0);
    \node[left] at (-0.15,0) {0};
    \draw (0,4) -- (-0.1,4);
    \node[left] at (-0.15,4) {1};

    \node[below] at (2, -0.5) {Difficulty};
    \node[rotate=90] at (-0.5, 2) {Probability};

    \draw[very thick] (0,0) rectangle (4,4);
    
    \draw[very thick, purple] (0,0) rectangle (2,4);
    
    \draw[very thick, orange] (0,0) rectangle (3,2);
    
    \begin{scope}[shift={(5,3.5)}]
      \draw[very thick, purple] (0,0) -- (0.8,0);
      \node[right] at (1,0) {Human-Solvable};
      
      \draw[very thick, orange] (0,-0.7) -- (0.8,-0.7);
      \node[right] at (1,-0.7) {AI-Solvable};
    \end{scope}

    \end{tikzpicture}
    \caption{AI as a Solver.}
    \label{solver}
\end{figure}
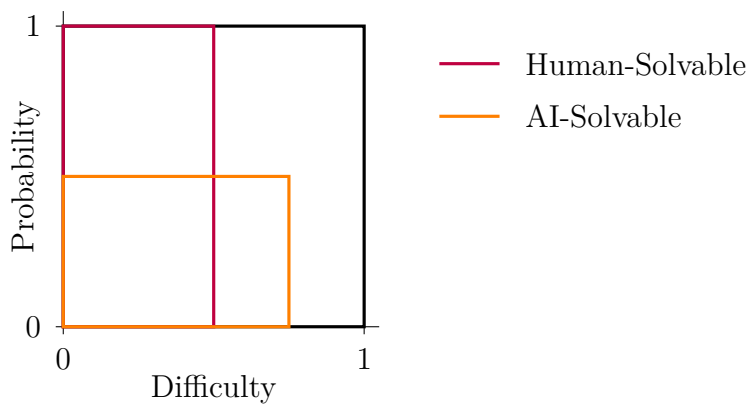

In this case, the set of problems AI can solve is \textit{not} a subset of the problems this human can solve. As such, for some problems (of difficulty $1/2$ to $3/4$) \textit{only} the AI can make progress and AI is used as an independent solver. On the other hand, if human ability is greater than $d$, the AI is never needed to independently solve any problems. This case is plotted in the following figure, where $a$ is now $7/8$ instead of $1/2$.

\newpage

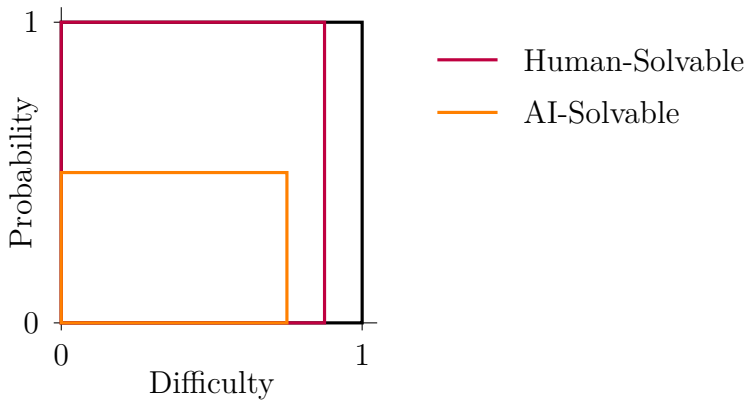
\begin{figure}[h]
    \centering
    \begin{tikzpicture}

    \draw (0,0) -- (4.2,0);
    \draw (0,0) -- (0,4.2);

    \draw (0,0.1) -- (0,-0.1);
    \node[below] at (0,-0.15) {0};
    \draw (4,0.1) -- (4,-0.1);
    \node[below] at (4,-0.15) {1};

    \draw (0,0) -- (-0.1,0);
    \node[left] at (-0.15,0) {0};
    \draw (0,4) -- (-0.1,4);
    \node[left] at (-0.15,4) {1};

    \node[below] at (2, -0.5) {Difficulty};
    \node[rotate=90] at (-0.5, 2) {Probability};

    \draw[very thick] (0,0) rectangle (4,4);
    
    \draw[very thick, purple] (0,0) rectangle (3.5,4);
    
    \draw[very thick, orange] (0,0) rectangle (3,2);
    
    \begin{scope}[shift={(5,3.5)}]
      \draw[very thick, purple] (0,0) -- (0.8,0);
      \node[right] at (1,0) {Human-Solvable};
      
      \draw[very thick, orange] (0,-0.7) -- (0.8,-0.7);
      \node[right] at (1,-0.7) {AI-Solvable};
    \end{scope}

    \end{tikzpicture}
    \caption{AI as a Helper.}
    \label{helper}
\end{figure}

Next, suppose humans simply choose to maximize the mass of problems they can solve with AI assistance.\footnote{This could correspond to a course with only homework and projects, where the student has full access to the AI technology. Additionally, this could correspond to a non-distortive measure of how problems are distributed in the world. If teachers could give assignments that prohibited the use of AI, that would not represent how problems can be solved outside of school. However, this is sometimes beneficial, as discussed in Section 7.} As problem difficulty is uniformly distributed\footnote{Re-labeling problems makes this without loss.} the mass of solvable problems is simply the area of the union of the two rectangles (human-solvable and AI-solvable). As such, humans who use AI as a solver solve:
$$a^s(t) \in \argmax_{a \in [0, d]} \left\{(1-p) a  + dp - c(a, t|d, p)\right\}$$
while humans who use AI as a helper solve: 
$$a^h(t) \in \argmax_{a \in [d, 1]} \left\{a - c(a, t|d, p).\right\}$$
Immediately, a key difference emerges. Humans who use AI as a solver have a marginal benefit of $1-p$ as learning more (at the margin) is only useful in cases where the AI hallucinates while humans who use AI as a helper have a marginal benefit of $1$ as problems on the margin for them are already too difficult for AI to tackle. This discontinuity in marginal benefit translates to a discontinuity in ability chosen by humans of different types. There is some threshold type $T$ for which humans of type below $T$ use AI as a solver while humans of type above $T$ use AI as a helper. Furthermore, all humans that use AI as a solver must have $a^s(t) \leq d$ while all humans that use AI as a helper must have $a^h(t) \geq d$. The following stylized plot displays $a^s(t)$ and $a^h(t)$ ignoring domain restrictions, as well as the induced ability mapping $A(t)$ that humans actually choose, along with the threshold type $T$.

\begin{figure}[h]
    \centering
    \begin{tikzpicture}
    
    \draw (0,0) -- (4.2,0);
    \draw (0,0) -- (0,4.2);
    
    \draw (0,0.1) -- (0,-0.1);
    \node[below] at (0,-0.15) {0};
    \draw (4,0.1) -- (4,-0.1);
    \node[below] at (4,-0.15) {1};
    
    \draw (0,0) -- (-0.1,0);
    \node[left] at (-0.15,0) {0};
    \draw (0,4) -- (-0.1,4);
    \node[left] at (-0.15,4) {1};

    \node[below] at (2, -0.5) {$t$};
    \node[left] at (0, 2) {$a$};
    
    \node[above] at (2, 4) {$T$};
    \node[left] at (4.5, 1.5) {$d$};
    
    \draw[very thick] (0,0) rectangle (4,4);
    
    \draw[very thick, blue] (0,0) -- (4,1.7);
    
    \draw[very thick, teal!80!black] (0,1) -- (4,4);
    
    \draw[dotted, thick] (0,1.5) -- (4,1.5); 
    \draw[dotted, thick] (2,0) -- (2,4); 

    \draw[very thick, orange] (0, 0) -- (2, 0.85);
    
    \draw[ultra thick, orange] (2, 2.5) -- (4, 4);

    \begin{scope}[shift={(5,3.5)}]
      \draw[very thick, teal!80!black] (0,0) -- (0.8,0);
      \node[right] at (1,0) {$a^h(t)$: AI as a Helper};
    
      \draw[very thick, blue] (0,-0.7) -- (0.8,-0.7);
      \node[right] at (1,-0.7) {$a^s(t)$: AI as a Solver};

      \draw[very thick, orange] (0,-1.4) -- (0.8,-1.4);
      \node[right] at (1,-1.4) {$A(t)$: Induced Ability Mapping};
    \end{scope}
    
    \end{tikzpicture}
    \caption{Ability Versus Type (Full)}
    \label{ability}
\end{figure}
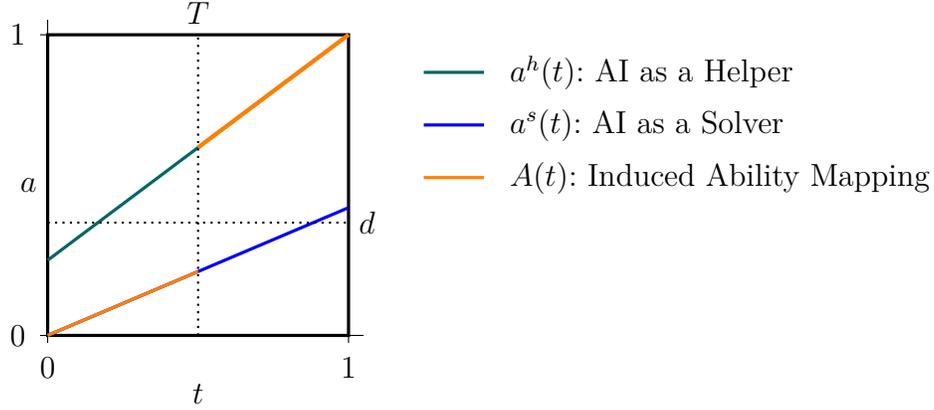

\begin{proposition}
  There exists a unique threshold type $T \in (0, 1)$ such that all humans of type $t < T$ use AI as a solver while all humans of type $t > T$ use AI as a helper. Then, $A(t) = a^s(t)$ for $t < T$ and $A(t) = a^h(t)$ for $t > T$. Furthermore, humans of type $T$ are indifferent between $a^s(T)$ and $a^h(T)$ and $A(t)$ has a discontinuity at $T$ with 
  $$\lim_{t \to T^-} A(t) = a^s(T) < d < a^h(T) = \lim_{t \to T^+} A(t).$$
\end{proposition}

\begin{proof}
  Clearly for all $t$:
  \begin{enumerate}
    \item $A(t) \in \{a^s(t), a^h(t)\}$;
    \item $A(t) = a^s(t)$ only if $a^s(t) \leq d$;
    \item $A(t) = a^h(t)$ only if $a^h(t) \geq d$.
  \end{enumerate}
  TThe only term in the human's objective function that depends on both $a$ and $t$ is the cost function $c(a, t|d, p)$. As such, decreasing differences in the cost function immediately implies that higher types choose weakly higher abilities. Thus, if $A(t) = a^s(t)$ for some $t$, then $A(t') = a^s(t')$ for all $t' > t$ as
  $$A(t') \geq A(t) = a^s(t) \geq d$$
  so $A(t')$ cannot be equal to $a^h(t')$. As such, a threshold $T$ exists (namely $T = \inf\{t: A(t) = a^h(t)\}$).

  Finally, indifference and the discontinuity. At $T$, humans must be indifferent between using AI as a solver or a helper as if not, continuity of payoffs implies that there would be some open set of humans around $T$ that would all choose either $A(t) = a^s(t)$ or $A(t) = a^h(t)$, contradicting the definition of $T$. If there were no discontinuity then $a^s(T) = a^h(T)$ and both would be equal to $d$ as $a^s(T) \leq d$ and $a^h(T) \geq d$. However, for investing slightly less to not be profitable, it must be that
  $$1-p > \frac{\partial}{\partial a} c(d, T|d, p)$$
  but for investing slightly more to not be profitable, it must be that
  $$1 < \frac{\partial}{\partial a} c(d, T|d, p)$$
  which is a contradiction as the above two inequalities imply $1-p > 1$.
\end{proof}

\section{Comparative Statics}

How do students' choice of ability vary as the AI technology varies? As the probability AI solvers are correct and the difficulty of problems they can solve varies, student incentives are impacted through changes in the returns to learning (i.e. the mass of problems students of a certain ability can solve that are otherwise not solvable by AI) and through changes in the cost of learning. These two forces impact how students who use AI as a helper versus solver decide their optimal level of ability, as well as the threshold determining how AI is used. If advances in AI technology generally increase its ability to help students learn, then every type conditional on using AI as a helper or a solver chooses a higher ability. Increasing the probability AI does not hallucinate decreases the marginal benefit of investment for individuals that use AI as a solver but does not change the marginal benefit for those that use AI as a helper. On the other hand, increasing the difficulty of problems AI can solve has no marginal impact on those that use it as a solver but decreases the incentive to switch to using AI as a helper instead of as a solver. We will now explore these forces.

\subsection{Ability and $p, d$}

Humans that use AI as a helper solve
$$a^h(t) \in \argmax_{a \in [d, 1]} \left\{a - c(a, t|d, p)\right\}.$$
Increasing differences in $a, p$ holds if $\frac{\partial^2 c}{\partial a \partial p} < 0$ and increasing differences in $a, d$ hold if $\frac{\partial^2 c}{\partial a \partial d} < 0$ so the reductions in learning costs when AI advances are larger the more advanced the human is.

On the other hand, humans that use AI as a solver solve
$$a^s(t) \in \argmax_{a \in [0, d]} \left\{(1-p) a  + dp - c(a, t|d, p)\right\}.$$
The cross partial of the objective with respect to $a, p$ is
$$-1 - \frac{\partial^2 c}{\partial a \partial p}$$
so increasing differences holds if
$$\frac{\partial^2 c}{\partial a \partial p} < -1.$$
Note that in this case, increases in $p$ also decreases the marginal value of increasing $a$, leading to the condition being harder to satisfy. Like before, increasing differences in $a, d$ hold if $\frac{\partial^2 c}{\partial a \partial d} < 0$.

This presents one way for advances in AI to increase inequality between students of different types: It is more difficult for advances in AI to incentive humans who use AI as a solver to invest in higher ability than it is for humans who use AI as a helper. Furthermore, as the ability of humans who use AI as a solver is more sensitive to changes in $p$ than humans who use AI as a helper are, increases in $p$ may increase the magnitude of the discontinuity between those that use AI as a solver and those that use it as a helper. As such, careful targeting of AI tools is crucial to prevent increases in inequality.

\subsection{Changes in the Solver-Helper Threshold}

Next, consider the threshold. Humans are indifferent between using AI as a solver or a helper at the threshold value. Let 
$$U^h(t| d, p) = a^h(t| d, p) - c(a^h(t| d, p), t| d, p)$$
and 
$$U^s(t| d, p) = (1-p)a^s(t| d, p) - c(a^s(t| d, p), t| d, p) + dp$$ 
be induced utilities so the threshold value $T = T(d, p)$ solves
$$U^h(T| d, p) = U^s(T| d, p).$$
Suppose $p$ increases. Then, 
$$\frac{\partial U^h(T| d, p)}{\partial p} = -\frac{\partial c(a^h(T| d, p), T| d, p)}{\partial p}$$
while 
$$\frac{\partial U^s(T| d, p)}{\partial p} = -a^s(T| d, p) - \frac{\partial c(a^s(T| d, p), T| d, p)}{\partial p} + d$$
and the threshold will tilt in favor of which utility grows more (as students who were previously indifferent at $T$ will shift to that side). In particular, $T$ grows when $p$ increases (at the margin) if and only if
$$-\frac{\partial c(a^h(T| d, p), T| d, p)}{\partial p} > -\frac{\partial c(a^s(T| d, p), T| d, p)}{\partial p} + (d-a^s(T| d, p))$$
which is equivalent to 
$$\int_{a^s(T| d, p)}^{a^h(T| d, p)} - \frac{\partial^2 c(a, T| d, p)}{\partial a\partial p} da > d-a^s(T| d, p)$$
by the Fundamental Theorem of Calculus. Similarly, $T$ grows when $d$ increases if and only if 
$$-\frac{\partial c(a^h(T| d, p), T| d, p)}{\partial d} > -\frac{\partial c(a^s(T| d, p), T| d, p)}{\partial d} + p.$$
which is equivalent to
$$\int_{a^s(T| d, p)}^{a^h(T| d, p)} - \frac{\partial^2 c(a, T| d, p)}{\partial a\partial d} da > p.$$
As such, complementarities between ability and models that can solve more difficult problems or solve problems with greater accuracies determines the direction the threshold moves. Increasing $d, p$ only benefits humans who use AI as a helper through reduced costs, but benefits humans who use AI as a solver both through reduced costs \textit{and} through the additional problems AI can solve independently. Since $d$ and $p$ are complementary in determining the mass of problems AI can solve, changes in the mass of additional problems that AI can solve independently as $d, p$ change are larger when the other of $p, d$ is larger. 

One consequence is that the more advanced AI is, the more difficult it is for advances in AI to \textit{increase} human learning: The higher $d, p$ are, the larger $-\frac{\partial^2 c}{\partial a \partial p}, -\frac{\partial^2 c}{\partial a \partial d}$ have to be to avoid the threshold increasing. This presents another possibility for increasing inequality: If the threshold were pushed extremely close to $1$ (i.e. almost all students use AI as a solver) due to advanced AI, only a small number of humans would have ability exceeding that of AI.

\subsection{Changes in Cost}

Finally, consider changes purely in the cost function (i.e. teacher training on how to best use the same underlying LLM). Suppose that the cost function $c$ is decreased by $k$ so costs to learning are $c(a, t) - k(a, t)$ (where dependency on $d, p$ is dropped). For both humans who use AI as a solver and humans who use AI as a helper, this change in costs only leads to an increase in the objective of $k(a, t)$. As such, humans with higher chosen levels of ability are helped more if $\partial k(a, t)/\partial a > 0$. Similar to before, if a reduction in learning costs is more significant when investing in high ability than low ability, then humans that use AI as a helper or as a solver all choose higher ability. This condition is also sufficient for the threshold to decrease, thereby increasing net human intelligence: When moving from cost $c$ to $c-k$, utility for humans at the threshold type $T$ that use AI as a helper increases by $k(a^h(T| d, p), T)$ while utility for humans of type $T$ who use AI as a solver increases by $k(a^s(T| d, p), T)$. Then, 
$$k(a^h(T| d, p), T) - k(a^s(T| d, p), T) = \int_{a^h(T| d, p)}^{a^s(T| d, p)} k'(a, T) da > 0$$
as $a^h(T| d, p) < a^s(T| d, p)$ so more humans use AI as a helper. The following are two natural forms of cost reductions that satisfy $k'(a, t) > 0$:
\begin{enumerate}
    \item Proportional Savings: Suppose $k(a, t) = (1-\lambda) c(a, t)$ for some $\lambda \in (0, 1)$. Then, $k' = (1-\lambda) c' > 0$ as $c' > 0$ and $c-k = \lambda c$.
    \item Cumulative Improvements: Suppose costs are cumulative so that a student who wishes to achieve ability $a$ must first incur all the costs to achieve ability $a' < a$. This is equivalent to working with $c(a, t) = \int_0^a c'(\alpha, t) d\alpha$. If reductions in costs are also cumulative, so a student wishing to achieve ability $a$ also benefits from the reductions in cost to achieve ability $a' < a$, then we can write $k(a, t) = \int_0^a Mk(\alpha, t) d\alpha$ where $Mk$ stands for the marginal reduction in costs at a given ability and type. Then, $k'(a, t) = Mk(a, t)$ is greater than zero as long as the marginal cost to learn anything is reduced.
\end{enumerate}

\section{Misspecified Students}

Fix an AI technology's $d, p$ and some cost function $c(a, t)$, suppressing dependence on $d, p$. It is often difficult to understand how often the AI is correct or if it hallucinates, leading to potential over-reliance. To model this, suppose students behave as if the AI is correct with probability $p' (> p)$.

On the other side, instructors also have flexibility in the type of assessments they can give. They can choose to assign homework, projects, or take-home/open-book exams that permit the usage of AI. Alternatively, they can give closed-book exams that do not permit the usage of AI. Suppose instructors put weight $\lambda$ on assignments that permit the use of AI and $1-\lambda$ on assignments that do not permit the use of AI. In this case, students who use AI as a solver solve
$$a^s_m(t) = \argmax_{a \in [0, d]} \left\{\lambda\big[(1-p')a+dp'\big] + (1-\lambda)a - c(a, t) \right\}$$
where the subscript $m$ denotes misspecification. Re-writing the objective gives
$$a^s_m(t) = \argmax_{a \in [0, d]} \left\{a + \lambda p'(d-a) \right\}.$$
Students that use AI as a helper have the same objective as before as they did not rely on AI even if it were (or were not) available on certain assignments:
$$a^h(t) = \argmax_{a \in [d, 1]} \{a - c(a, t)\}.$$

Taking $\lambda = p/p'$ leads to the misspecified student's problem coinciding with the true problem. When $p = p'$ and there is no gap between how sophisticated students think AI is and how sophisticated it actually is, the optimal instructor policy is to allow access to AI on all assignments (to best align with humans having access to AI tools day-to-day).

Educators have already begun to shift in this direction: In response to too many AI generated research papers, one Northeastern professor indicated that 

\begin{quote}
  I will redesign the assignment so it can't be done with AI next time. I had one student complain that the weekly homework was hard to do and they were annoyed because Claude and ChatGPT was useless in completing the work. I told them that was a compliment, and I will endeavor to hear that more from students \citep{b_2025_anthropic}.
\end{quote}

\section{Discussion}

I consider a simple model featuring two stylized facts about AI: (1) it hallucinates with some probability, even when engaging in material it should ``know'' and (2) it has emergent capabilities, where certain abilities appear suddenly as model size and training data increase, leading to sharp cutoffs in what AI can accomplish. Students have access to this technology, which both has the ability to autonomously solve problems and help students learn at a lower cost. Students choose their optimal ability to maximize the mass of problems they can solve (with AI assistance), net of the cost of learning. When faced with this problem, students separate into two groups: Those that use AI as a solver have lower ability than AI and rely on it to solve some problems independently, while those that use AI as a helper have higher ability than AI and only use it to help them learn at a lower cost.

This model makes several predictions. First, advances in AI technology that increase the probability AI does not hallucinate or the difficulty of problems it can solve do not always lead to increases in human learning, nor does it even necessarily increase the total amount of problems a human and an AI can jointly solve. In general, there is a tradeoff between AI which hallucinates less and the marginal benefit for humans that use AI as solvers to learn more. Additionally, the tradeoff between human and AI intelligence leads to increases in inequality between students of different types: Students that use AI as a helper have a stronger incentive to learn compared to students who use AI as a solver do. In particular, there is a discontinuity in ability between students who use AI as a solver and students who use AI as a helper, and this gap may increase as AI technology advances. Finally, changing educational tools is one way to remedy the potential negative effects of AI on human learning. If all school assignments prohibited the usage of AI, the discontinuity would disappear (as students cannot use AI as an independent solver), but would not be representative of the real world. However, prohibiting use on some portion of assignments may counteract misspecified beliefs about how accurate AI actually is, restoring efficiency.

Many open questions remain. Most interesting would be to take the model to data. Different models and domains have different values of $d, p$ which could influence how students use AI tools. Students may directly ask for solutions more often in foundational classes, characterized by higher $d, p$, than in more advanced courses.\footnote{Of course, self-selection into these classes may be another confounding variable explaining different usage patterns.}  Next, the model considered in this paper is extremely simplistic. There is only one dimension of problems but students often require multiple complementary skills. Whether different areas are complements or substitutes may have qualitative impacts on how human behavior and ability looks like in a world with AI. Additionally, the payoff function is extremely simple while in reality, students may make more nuanced decisions about when and how to use AI tools. Even humans that use AI as a helper may sometimes use AI to independently solve parts of a problem (such as doing routine calculations) or to check their work, providing an avenue for AI to help humans that know more than it. Different students may also have different amounts of time available to them, once again influencing their demand for AI assistance. Understanding how teachers should best design materials facing heterogeneity in student circumstances is of importance. This paper only considered the simple one-dimensional parametrization of misspecification a single student has; different students may be misspecified to different degrees, which makes determining the optimal level of assignments without AI a more difficult problem. Furthermore, the distribution of problems in a class will be quite different from the distribution of problems in the real world, and a student may derive utility from both a good grade and learning for downstream tasks (like passing a job interview). Finally, AI and human learning is a two way street: Just like how humans learn from AI, AI can also learn from humans who can provide feedback on output and new training data. As such, having too few humans with ability exceeding that of AI may lead to stagnation in AI progress as well. 

\newpage

\bibliography{cites}

\end{document}